\documentclass[aps,prd,a4paper,twocolumn,showpacs,nofootinbib]{revtex4}
\usepackage{amssymb,amsmath,latexsym}
\usepackage{graphicx}
\usepackage{epsfig}
\usepackage{varioref,xr-hyper}
\usepackage{color}

\usepackage[hypertex,draft=false,verbose=false,debug=false
            hyperindex=true,hypertexnames=true]{hyperref}

\newcommand{\hide}[1]{{}}

\newcommand{\be}{\begin{equation}}
\newcommand{\ee}{\end{equation}}
\newcommand{\bea}{\begin{eqnarray}}
\newcommand{\eea}{\end{eqnarray}}

\newcommand{\begm}{\begin{pmatrix}}
\newcommand{\enm}{\end{pmatrix}}

\def\n{{\bf  \hat n}}
\def\k{{\bf k}}

\def\lsim{\;\raise 0.4ex\hbox{$<$}\kern -0.8em\lower 0.62 ex\hbox{$\sim$}\;}
\def\gsim{\;\raise 0.4ex\hbox{$>$}\kern -0.7em\lower 0.62 ex\hbox{$\sim$}\;}

\begin{document}

\title{From Cavendish to PLANCK: Constraining Newton's Gravitational Constant with
CMB Temperature and Polarization Anisotropy}

\author{Silvia Galli}
\email{galli@apc.univ-paris7.fr}
\affiliation{Physics Department, Universita' di Roma ``La Sapienza'',
  Ple Aldo Moro 2, 00185, Rome, Italy\\
Laboratoire Astroparticule et Cosmologie (APC), Universite' Paris Diderot - 75205 PARIS cedex 13.}

\author{Alessandro Melchiorri}
\email{alessandro.melchiorri@roma1.infn.it}
\affiliation{Physics Department and INFN, Universita' di Roma ``La
  Sapienza'', Ple Aldo Moro 2, 00185, Rome, Italy}

\author{George F. Smoot}
\email{gfsmoot@lbl.gov}
\affiliation{Lawrence Berkeley National Laboratory and
Berkeley Center for Cosmological Physics Physics
  Department, University of California, Berkeley CA 94720}

\author{Oliver Zahn}
\email{zahn@berkeley.edu}
\affiliation{Lawrence Berkeley National Laboratory and
Berkeley Center for Cosmological Physics Physics
  Department, University of California, Berkeley CA 94720}

\begin{abstract}
\noindent
We present new constraints on cosmic variations of Newton's gravitational constant by making use of the latest
CMB data from WMAP, BOOMERANG, CBI and ACBAR experiments and independent constraints coming from
Big Bang Nucleosynthesis. We found that current CMB data provide constraints at the 
$\sim 10 \%$ level, that can be improved to $\sim 3 \%$ by including BBN data.
We show that future data expected from the Planck satellite could constrain 
$G$ at the $\sim 1.5 \%$ level while an ultimate, cosmic variance limited, CMB experiment 
could reach a precision of about $0.4 \%$, competitive with current laboratory measurements.
\end{abstract}


\maketitle

\section{Introduction}

Since Cavendish's first measurement in $1798$ (\cite{cavendish}), Newton's Gravitational constant remains one of the most elusive constants in physics. The past two decades did not succeed in substantially improving our knowledge of its value
from the precision of $0.05 \%$ reached in $1942$ (see \cite{1942}). To the contrary, the variation between different measurements forced the CODATA committee\footnote{See http://www.codata.org/}, which determines the internationally accepted standard values, to increase the uncertainty from
$0.013 \%$ for the value quoted in $1987$ to the one order of magnitude larger uncertainty of $0.15 \%$ for the $1998$ "official" value (\cite{codata}). Recent laboratory measurements (see e.g. \cite{gother}) point towards an uncertainty at the level of $\sim 0.4 \%$, while other works claim an improved precisions below $0.01 \%$ (\cite{claims}).  Analysis of the secular variation of the period of nonradial pulsations of the white dwarf G117-B15A (\cite{gwd}) has produced complementary constraints at $\sim 0.1 \%$ level.

\noindent Measurements of the Cosmic Microwave Background (CMB, hereafter) temperature and polarization anisotropy have been suggested as a possible tool for determining the value of $G$ (see \cite{zalzahn}).
In recent years, CMB temperature and polarization anisotropy have been measured with great precision from experiments as WMAP (\cite{wmap5cosm,wmap5komatsu}), BOOMERANG (\cite{boom03}), CBI \cite{cbi} and ACBAR (\cite{acbar}). The impressive agreement between those measurements and the expectations of the standard model of structure formation have paved the way to the use of cosmology
as a new laboratory where to test physical hypothesis at energies and scales not reachable on earth.
Since a variation in $G$ affects CMB temperature and polarization anisotropy, changing the position and the amplitude of the acoustic peaks present in the corresponding angular power spectra, it is indeed possible to infer new and independent constraints on $G$
from CMB data.

In this paper we follow this timely line of investigation. Respect to previous works (most notably \cite{zalzahn}) we update the
CMB constraints on $G$ by using the most recent CMB data (most notably, WMAP) and by also including  
complementary information from Big Bang Nucleosynthesis (hereafter, BBN, see \cite{Iocco:2008va} for a complete review).
As already shown in several papers (see e.g. \cite{bbng2}, \cite{barrow}), any variation in $G$ changes the Hubble parameter at BBN given
by $H \sim {\sqrt{Gg_*}} T^2$ where $g_*$ counts the number of relativistic particles species and $T$ is the temperature of
the Universe. Since the predicted amount of light elements depends crucially on the comparison between the expansion rate
$H$ and, for example, the neutron-proton conversion rate $\Gamma_{\rm np}\sim G_F^2T^5$, where $G_{F}$ is the Fermi constant,
any change in $G$ can be strongly constrained by combining BBN predictions with observations of primordial elements.
Moreover, we also discuss the ability of next CMB experiments as Planck (\cite{planck}) to constrain $G$, including the possibility of a  "cosmic variance limited" survey. 

Any cosmological constraint is, however, indirect and, in the case of the CMB data, depends on the assumed theory of structure formation.
The major caveat in our case is the assumption of a cosmological constant, or dark energy component, the nature of which is puzzling
and unknown (for a recent review, see e.g. \cite{silvestri}, \cite{copeland}, \cite{ratra}).
While the derived constraints will therefore be model dependent, it is interesting that a major alternative to
a dark energy component, i.e. modified gravity theories, could be parameterised by introducing an
effective value of Newton's constant $G_{\rm eff}$, that could not only be different from the local value
of $G$ but also spatial and time dependent (see e.g. \cite{daniel}, \cite{berti},\cite{bransdicke}).
Moreover, if dark energy interacts with dark matter, there is a change in the background evolution of the
universe leading to an effective $G_{\rm eff}$ for the matter component (see e.g. \cite{marteens})
and to a possible change in the cosmic bound on $G$.

In this respect, the search for variations in Newton's constant
using cosmological data could also play a role in the understanding of the dark sector. If the
Newton's constant inferred from cosmology will turn out to be different from the local value, then 
this may suggest a modification of gravity at large scale or a more complex interacting 
dark energy scenario. Since an interacting dark energy or a modified gravity theory could be responsible for a
variation of $G$ in the late universe, we also consider the possibility of a redshift dependence of $G$.

Our paper is therefore organized as follows: in the next section we briefly describe the effects of a variation
in $G$ on CMB temperature and polarization anisotropy. In Section III we describe our method of analysis and the
datasets considered. In Section IV we present our results and, finally, in Section V we derive our conclusions.

\section{The Impact of $G$ on recombination and the CMB}

Following \cite{zalzahn} we parameterize the deviations from Newton's gravitational
constant by introducing a dimensionless parameter $\lambda_G$ such that
\begin{equation}
G\rightarrow \lambda_G^2 G
\end{equation}

As showed in \cite{zalzahn}, expressing the perturbed quantities in Fourier space,
a variation in Newton's gravitational constant is
equivalent in a simple 
re-scaling of the wave numbers. No preferred cosmological scale is introduced by varying $G$ 
and the density fluctuations produced by a mode of
wavevector $\k$ in a universe with $\lambda_G \neq 1$ have equivalent dynamics 
of a mode with $k'=k/\lambda_G$ in a universe with $\lambda_G= 1$.

\begin{figure}[h!]
 \includegraphics[width=0.4\textwidth]{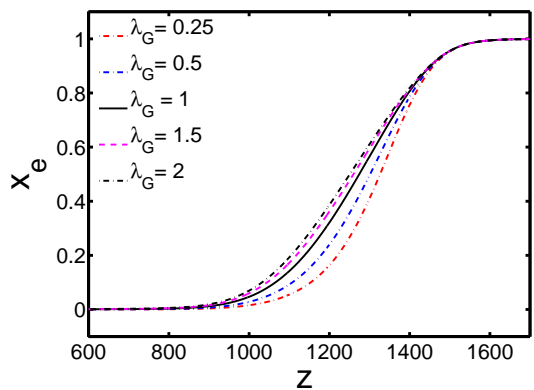}
\caption{Ionization fraction in function of redshift for different values of
$\lambda_G$.}
\label{iong}
\end{figure}

However the physics of recombination does introduce a preferred timescale and it
 will actually change when varying $\lambda_G$.
This is clearly shown in Figure \ref{iong} where
the ionization fraction $x_e$  at different redshift $z$, computed with a modified version of
RECFAST \cite{recfast}, is plotted for different
values of $\lambda_G$. The ionization fraction $x_e$ is just the free electron number density $n_e$ divided by the total number density of hydrogen nuclei (free and bound) $n_H$. As we can see, higher (lower) values of the gravitational constant
yields a delayed (accelerated) period of recombination.
A change in the number density of free electrons $n_e$ 
in function of the conformal time $\tau$, changes the visibility function $g(\tau)$, 
written in terms of the opacity for Thomson scattering $\kappa$ as
\begin{equation}
g(\tau)=\dot{\kappa} \exp(-\kappa)=-d/d\tau \exp(-\kappa)
\end{equation}
with
\begin{equation}
\kappa = \sigma_T \int_\tau^{\tau_0} a n_e(\tau) d\tau,
\label{opacity}
\end{equation}
\noindent where $\sigma_T$ is the Thomson scattering cross section, $a$ is the scale
factor and  $\dot{\kappa}=\sigma_T a n_e $.

This clearly affects the CMB temperature anisotropy that can be written as an 
integral along the line of sight over sources,

\begin{equation}
\Delta T(\n,\k) = \int^{\tau_0}_0  d \tau \;  S(k,\tau) e^{i \k\cdot \n D(\tau)}  g(\tau)
\label{lineofsight}
\end{equation}

\noindent where $S(k,\tau)$ is the anisotropy source term (see \cite{Seljak:1996is})
and $D(\tau)$ is the distance from the observer to a point along the line of sight at conformal time 
$\tau$.

\begin{figure}[h!]
 \includegraphics[width=0.4\textwidth]{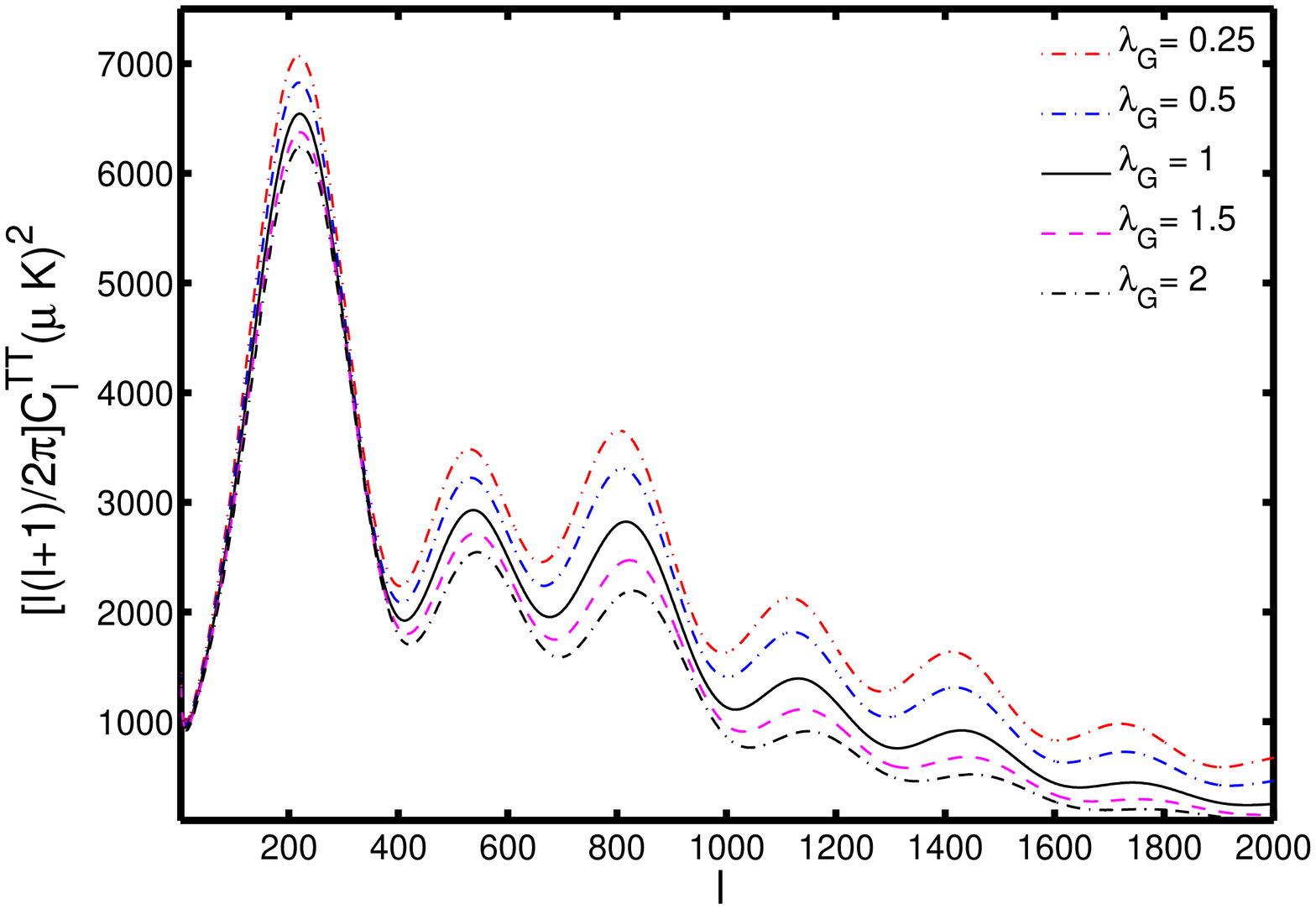}
 \includegraphics[width=0.4\textwidth]{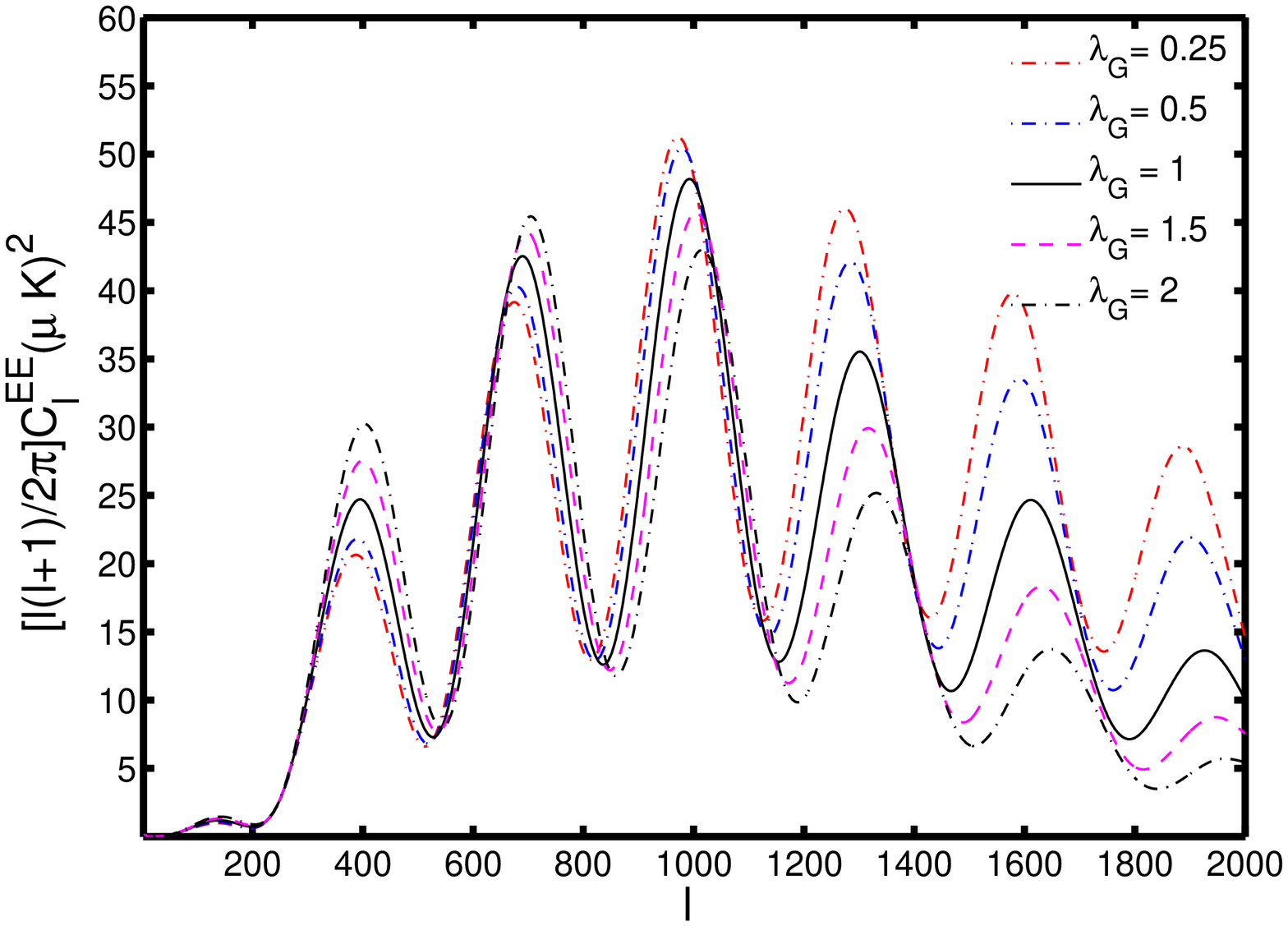}
 \includegraphics[width=0.4\textwidth]{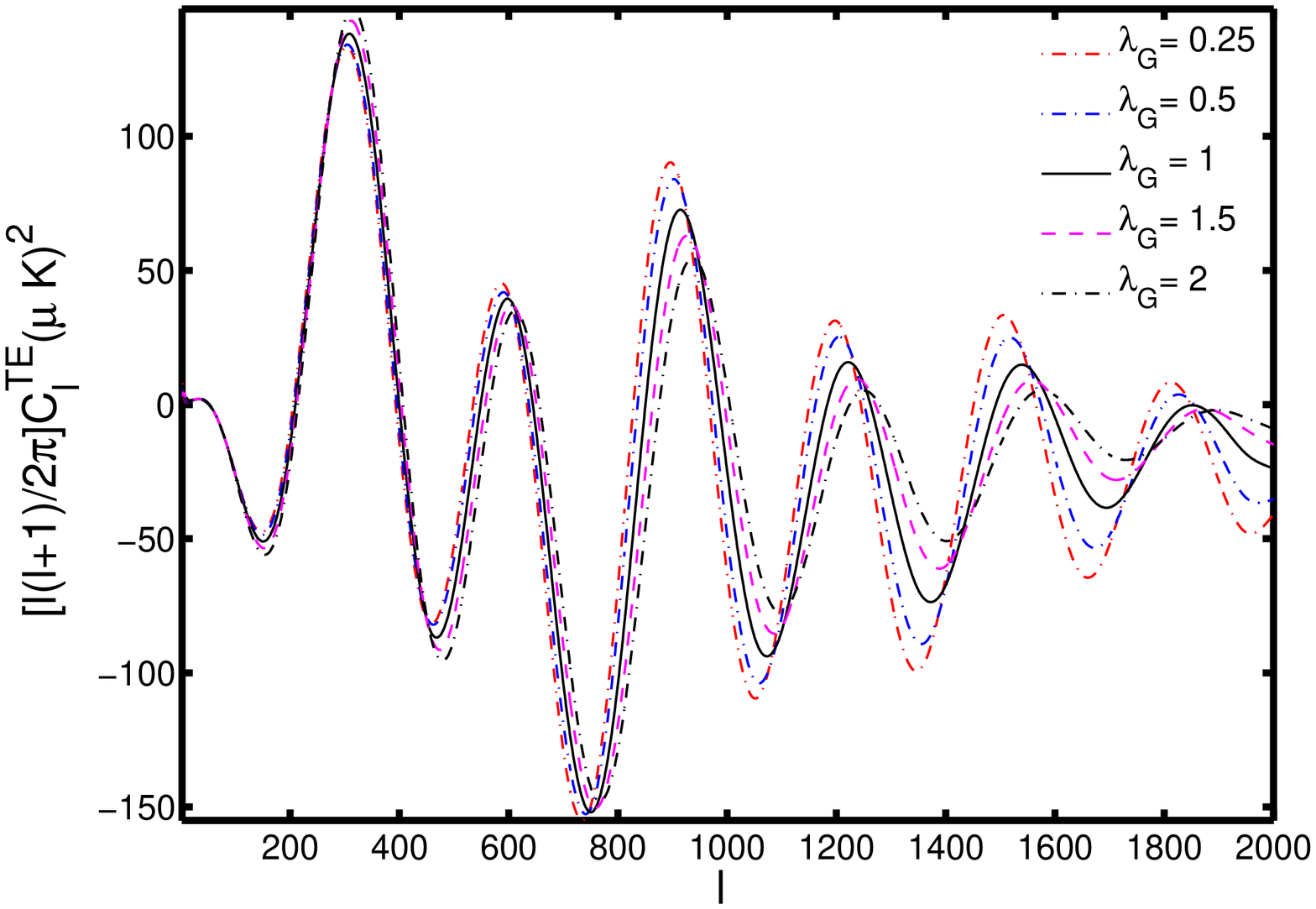}
\caption{From Top to Bottom: Temperature, Polarization and cross
Temperature-Polarization power spectra in function of
variations in $\lambda_G$.}
\label{spettri}
\end{figure}

In Figure \ref{spettri} we plot the CMB temperature and polarization
spectra computed from a modified version of the CAMB \cite{camb} code.
The effect of modified recombination is clear. Namely, varying $\lambda_G$ changes
the recombination process, shifting $g(\tau)$
along the conformal time $\tau$. The net effect is a damping or enhancement 
of the acoustic oscillations and a shift of the Doppler peaks in the angular scales. 
This mechanism could mimic an extra injection or absorption of
Lyman-$\alpha$ photons at last scattering, as already analyzed
in several recent papers (see e.g. \cite{recmod}),
and it would be difficult to disentangle the two scenarios.

Another important aspect to consider is a possible redshift dependence of
$G$. If interacting dark energy or a modification to general relativity are responsible
for the current accelerated expansion of the universe, it is indeed
possible that this could result in an observed cosmic value of $G$ different
from the one obtained from local measurements. Moreover, it is plausible to
think that this kind of deviation of $G$ will be triggered by
acceleration, i.e., to be conservative, will appear at redshift $0.1 < z < 2$.

We have therefore considered two possible parameterizations for a redshift-dependent
gravitational constant.
A first parameterization, that somewhat ties the change in $G$ with the appearance of
dark energy is to consider:

\begin{equation}
G(z)=G+\Delta G (1-a)
\label{smoot}
\end{equation}
where the variation $\Delta G$ is equal to $G(\lambda_G^2-1)$. This parameterization, similar to the one proposed in
\cite{CPL} for the dark energy equation of state, has the advantage of a smooth transition between
the value of $G$ today to $\lambda_G^2 G$ in the past, when $z \gg 1$. However the
redshift of transition between these two values is not an independent variable.

We have therefore considered a second possible parameterization as:

\begin{equation}
G(z)=G[1-(1-\lambda_G^2)H(z-z_t)]
\label{heavi}
\end{equation}

\noindent where $H(x)$ is the Heaviside function ($H(x)$ for $x<0$ and
$H(x)=1$ for $x>0$) and $z_T$ is the redshift of transition between
the two values (local and past) of $G$.

\begin{figure}[h!]
 \includegraphics[width=0.4\textwidth]{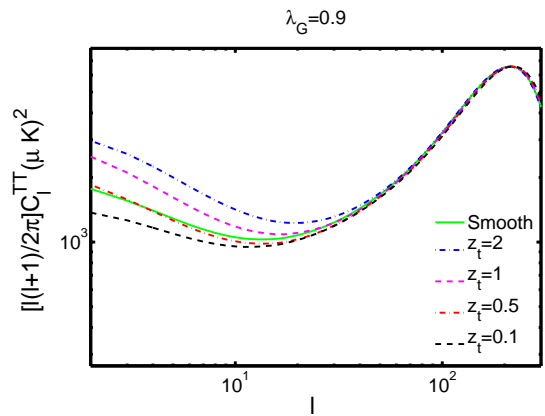}
\caption{Temperature
power spectrum when the gravitational constant varies such that
$\lambda_G =0.9 $. The graph shows the effects of the smooth transition parameterization described in equation \ref{smoot} (green line) and of the Heaviside parameterization of equation \ref{heavi} for different redshifts of transition $z_T$ between 0.1 and 2.}
\label{redshift}
\end{figure}

In Figure \ref{redshift} we plot different power spectra computed considering the two 
parameterizations described using a fixed value of $\lambda_G=0.9$. As we can see,
introducing a redshift dependent variation in $G$ increases the CMB anisotropy
at large angular scales. On sub-Hubble scales, the Einstein equations in an expanding space-time
reduce to the Poisson equation

\begin{equation}
 \Delta\Phi = 4\pi G\rho a^2 \delta
\end{equation}

\noindent that relates the gravitational potential $\Phi$ to the density contrast $\delta$.
If a redshift variation in $G$ occurs, this will clearly change the
gravitational potential, the density growth function and large scale CMB anisotropy through
the Integrated Sachs Wolfe effect (ISW hereafter, see e.g. \cite{hu}).
Since a large ISW signal is at odds with current WMAP data, a varying with
redshift $G$ is strongly constrained, as we will see in the next section.

\section{Analysis Method}

We constrain variations in the Newton's constant
with current CMB data by making use
of the publicly available Markov Chain Monte Carlo
package \texttt{cosmomc} \cite{Lewis:2002ah}.
Other than $\lambda_G$ we sample the following set of cosmological parameters,
adopting flat priors on them:
the physical baryon and CDM densities, $\omega_b=\Omega_bh^2$ and
$\omega_c=\Omega_ch^2$, the Hubble parameter, $H_0$, the scalar
spectral index, $n_{s}$,
the normalization, $\ln10^{10}A_s(k=0.05/Mpc)$ and the reionization
optical depth $\tau$.

As discussed in the previous section, we will also consider the possibility
of a variation with redshift in $G$ and we will consider as extra parameter
the redshift of transition $z_T$.

The MCMC convergence diagnostic tests are performed on $4$ chains using the
Gelman and Rubin ``variance of chain mean''$/$``mean of chain variances'' $R-1$
statistic for each parameter. Our $1-D$ and $2-D$ constraints are obtained
after marginalization over the remaining ``nuisance'' parameters, again using
the programs included in the \texttt{cosmomc} package.
We use a cosmic age top-hat prior as 10 Gyr $ \le t_0 \le$ 20 Gyr.
We include the five-year WMAP data \cite{wmap5komatsu} (temperature
and polarization) with the routine for computing the likelihood
supplied by the WMAP team (we will refer to this analysis as WMAP5).

Moreover, in order to test the effect of current polarization
measurements on constraining $\lambda_G$ we also considered
the combination of the WMAP data with the polarization results
coming from the BOOMERANG (\cite{boom03}) and CBI (\cite{cbi})
experiments. We will refer to this analysis as WMAP5+POL.

Together with the WMAP data we also
consider the small-scale CMB measurements of ACBAR
\cite{acbar} (we will refer to this analysis as WMAP5+ACBAR).

Finally, we forecast future constraints on $\lambda_G$
simulating a set of mock data with a fiducial model
given by the best fit WMAP5 model with $\lambda_G=1$
and experimental noise described by:
\begin{equation}
N_{\ell} = \left(\frac{w^{-1/2}}{\mu{\rm K\mbox{-}rad}}\right)^2
\exp\left[\frac{\ell(\ell+1)(\theta_{\rm FWHM}/{\rm rad})^2}{8\ln 2}\right],
\end{equation}
\noindent where $w^{-1/2}$ is the temperature noise level
(we consider a factor $\sqrt{2}$ larger for polarization
noise) and  $\theta$ is the beam size.
We considered two future datasets. The first,
based on the experimental specifications of the PLANCK
SURVEYOR mission, with $w^{1/2}=58 \mu K$ and $\theta_{\rm FWHM}=7.1'$
equivalent to the $143$ GHz channel (see \cite{planck}).
The second dataset is a cosmic variance
limited experiment (CVL hereafter) with no experimental
noise for both temperature and polarization anisotropy and 
$\ell_{\rm max}=2500$.

Constraints on $\lambda_G$ are also computed using
standard BBN theoretical predictions as provided by the new
numerical code described in \cite{Serpico:2004gx}\cite{Pisanti:2007hk}, which includes
a full updating of all rates entering the
nuclear chain based on the most recent experimental results on
nuclear cross sections. The BBN predictions are compared with
the D/H abundance ratio of \cite{O'Meara:2006mj} obtained including a new
measurement in a metal poor damped Lyman-$\alpha$ system along the
line of sight of QSO SDSS1558-0031 \be
\textrm{D/H}=(2.82_{-0.25}^{+0.27}) \cdot 10^{-5} \label{deut}
\ee We use the uncertainty as quoted in \cite{O'Meara:2006mj},
computed by a jackknife analysis.

\section{Results}

\subsection{Constant $G$ with redshift}

\begin{table}[b]
\begin{center}
\begin{tabular}{rc}
Experiment & Constraints on $\lambda_G$ at $68 \%$ c.l.\\
\hline
WMAP& $1.01\pm0.16$\\
\hline
WMAP+POL& $0.97\pm0.13$\\
\hline
WMAP+ACBAR& $1.03\pm0.11$\\
\hline
WMAP+BBN& $0.98\pm0.03$\\
\hline\hline
PLANCK& $1.01\pm0.015$\\
\hline
CVL& $1.002\pm0.004$\\
\hline
\end{tabular}
\caption{Constraints on $\lambda_G$ from current WMAP and BBN observations and future constraints
achievable from the Planck satellite mission and from a cosmic variance limited experiment.}
\label{constraints}
\end{center}
\end{table}

We report in Table \ref{constraints} the constraints obtained on $\lambda_G$ analyzing the datasets
mentioned in the previous section.
As we can see, current CMB data only provide a constraints at about $\sim 15 \%$ level.
The WMAP constraint is improved by $\sim 10 \%$ when temperature and polarization anisotropy data from BOOMERANG and
 CBI is included and by $\sim 30\%$ when the small scale temperature angular spectrum data from ACBAR is added.
However, as we can see from the Table, the major improvement comes from BBN:
in this case the constraint WMAP+BBN reaches the $\sim 3 \%$ level.

\begin{figure}[h!]
\includegraphics[width=0.4\textwidth]{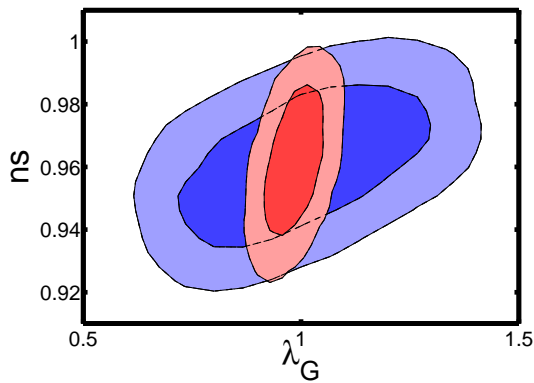}
\caption{$68 \%$ and $95 \%$ likelihood contour plots
on the $\lambda_G$-$n_S$ plane using present CMB data with and without BBN constraints.}
\label{nsconstraints}
\end{figure}

\begin{figure}[h!]
\includegraphics[width=0.4\textwidth]{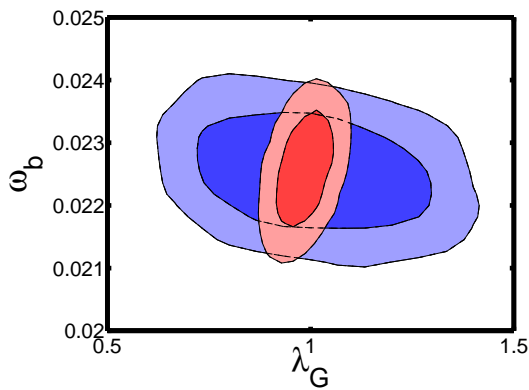}
\caption{$68 \%$ and $95 \%$ likelihood contour plots
on the $\lambda_G$-$\omega_b$ plane using present CMB data with and without BBN constraints.}
\label{obconstraints}
\end{figure}

It is interesting to consider possible correlations between $\lambda_G$ and more
usual cosmological parameters. In Figures \ref{nsconstraints} and \ref{obconstraints}
 we plot the $1$ and $2$ $\sigma$'s confidence level on the
$n_S$-$\lambda_G$ and $\omega_b$-$\lambda_G$ planes respectively.
As we can see there is a strong degeneracy between these parameters.
Increasing (decreasing) $G$ would yield higher (lower) values of $n_S$ and
lower (higher) values for $\omega_b$ more consistent with CMB data.

The degeneracy with the scalar spectral index is clear since increasing
$\lambda_G$ delays recombination, damping the small angular scale oscillations.
This effect could be counterbalanced by increasing $n_S$ and the small scale
power of primordial perturbations. This will also change the relative
amplitude between odd and even peaks, affecting the constraints on
the baryon density.

As already described in \cite{zalzahn}, another possible degeneracy is present
with the running of the spectral index $\alpha_s$. We have therefore
considered an extra analysis including possible variations in $\alpha_s$.
Considering the WMAP data only we found $\lambda_G=0.96\pm0.19$.

As we can see from Table \ref{constraints}, future experiments can substantially
improve the current constraints on $\lambda_G$.  The PLANCK Surveyor mission is 
expected to provide constraints at the $\sim 1.5 \%$ level. As already discussed in \cite{zalzahn},
the inclusion of polarization data is crucial in breaking the degeneracy between
$\lambda_G$ and inflationary parameters as $n_s$ and $\alpha_s$; we found that neglecting polarization data from Planck yields weaker constrains by a factor of $\sim 4$. Polarization data are therefore extremely useful in constraining $\lambda_G$.

The ultimate constraint achievable by a cosmic variance limited experiment is $0.4 \%$, competitive
with current laboratory bounds.

\subsection{Varying $G$ with redshift}

\begin{figure}[h!]
\includegraphics[width=0.4\textwidth]{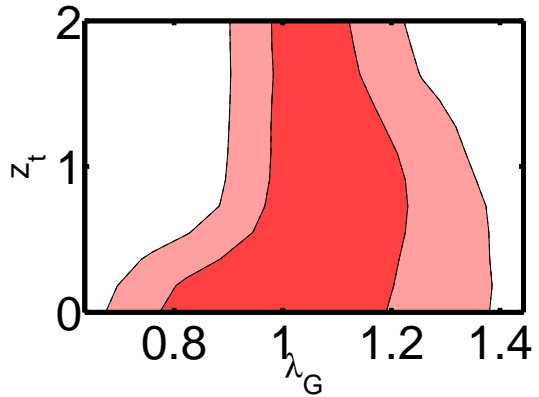}
\caption{$68 \%$ and $95 \%$ likelihood contour plots
on the $\lambda_G$-$z_T$ plane using present CMB data.}
\label{ztconstraints}
\end{figure}

Here we consider possible constraints on $G$ allowing for variations
in redshift. Using the simple parameterization in Equation \ref{smoot}
we found that the WMAP data alone yields the constraint
$\lambda_G=1.01\pm0.1$ at $68 \%$ c.l.. This constraint is better
by $\sim 40 \%$ respect the corresponding bound obtained with
constant $G$. The reason is due to the extra ISW effect that increases
the large angular scale CMB spectra, in disagreement with the
WMAP observations.

We have then considered a redshift dependence as in Equation \ref{heavi}
with a flat prior $0< z_T< 2$. In Figure \ref{ztconstraints} we plot
the $68 \%$ and $95 \%$ confidence levels on the $\lambda_G$-$z_T$ plane using
only the WMAP data. As we can see, for larger values of $z_T$ the constraints
on $\lambda_G$ are stronger. Again, the presence of the ISW effect,
irrelevant for $z_T\sim0$ but sizable for larger values,
helps in constraining $\lambda_G$.

\section{Conclusions}

In this paper we have updated the constraints from current CMB data on
Newton's gravitational constant $G$. We have found no evidence for
variation in this constant with a constraint of $\lambda_G=1.03\pm0.11$ at $68 \%$ c.l. 
from WMAP+ACBAR ($\lambda_G=0.98\pm0.03$ when BBN data is considered).
BBN plays therefore a crucial role in constraining $G$. 
However, even without considering the possibility of systematics in current
observations of primordial elements, the BBN constraints relies on the
perfect knowledge of the amount of relativistic degrees of freedom $g_*$.
Since $g_*=5.5+{7 \over 4} N_\nu^{eff}$ any possible extra background of
relativistic particles, parameterized by the effective number of neutrino
species $N_\nu^{eff}$ would drastically change the BBN bound.
Moreover, CMB and BBN probe completely different physics and epochs.
While the agreement between the two results is reassuring, it is clear
that it would be preferable to have an improved and independent CMB constraint.

We have then considered the constraints achievable from ongoing and future
satellite experiments. For the Planck Surveyor satellite mission
we have found a future constraints of the order of $1.5 \%$ using only
CMB data. Next, cosmic variance limited experiments as, for example,
 the future EPIC satellite proposal (see \cite{baumann}),
could probe Newton's constant with a $\sim 0.4 \%$ precision, i.e. with
grossly the same accuracy currently reached from local experiments.

It is important to stress that the accuracy on $\lambda_G$ achievable 
by the CMB is limited by 
how precisely we treat the recombination process.
Current recombination codes should be accurate enough for the
Planck mission (see e.g. \cite{rico}) but this may provide
an intrinsic limit for the next, beyond Planck, CMB surveys.
Moreover, recombination could be modified by non-standard mechanisms
as dark matter decay or variations in the fine structure constant $\alpha$. 
High frequency measurements of the black-body CMB spectrum, where 
recombination absorption lines are expected, could be helpful 
in disentangling the two effects. However, galactic foregrounds 
at those frequencies largely dominate over the CMB signal.

In this paper we followed a conservative approach by considering
only future CMB data. It is clear that the inclusion of complementary cosmological data, 
as expected from future galaxy, weak lensing and $21$cm surveys, will further break the degeneracies 
between the parameters and substantially improve
the constraints. We plan to discuss this in more detail in a future paper (\cite{galli2}).

Finally, we have considered a variation of $G$ with redshift, parameterizing its variation
either with a smooth transition between $G$ and $\lambda_G^2G$, or
with a simple step function at a transition redshift $z_T$.
The ISW effect arising from redshift variations in $G$ is at odds with the low
CMB quadrupole measured by WMAP and therefore yields stronger constraints on $\lambda_G$.
Current data, also in this case, do not exhibit a deviation from the standard value.
In this respect, the constraints obtained under the assumption of constant 
with redshift $G$ could be considered as more conservative.

Future cosmological data will therefore substantially improve the bounds on $G$ 
and on its possible variations with time, space and redshift. By comparing local and cosmic 
measurements, the Newton's constant will be less elusive and may shed light on the late accelerated evolution of the universe.

\vspace{0.6cm}

\noindent {\bf Acknowledgements}\\
The authors acknowledge helpful discussions with Ned Wright, John Barrow, Eric Linder and Fengquan Wu.
OZ acknowledges financial support of the Berkeley Center for Cosmological Physics.  
SG is supported by the \'Ecole Doctorale de Astronomie et Astrophysique de l'\^Ile de France.
This research has been supported by ASI contract I/016/07/0 "COFIS".

\end{document}